\documentstyle[12pt,dina4,ams,epsf]{article}
\newcommand{\be}{\begin{equation}\label}
\newcommand{\ee}{\end{equation}}
\newcommand{\bea}{\begin{eqnarray}\label}
\newcommand{\eea}{\end{eqnarray}}
\newcommand{\eps}{\epsilon}
\newcommand{\sgn}{\mbox{sgn}}
\begin{document}
%
%
\title{On the Collective Motion\\ in Globally Coupled Chaotic Systems}
\author{Wolfram Just\\
        Theoretische Festk\"orperphysik\\
        Technische Hochschule Darmstadt\\
        Hochschulstra\ss e 8\\
        D--64289 Darmstadt\\
        Germany}
\date{June 17, 1996}
\maketitle
\begin{abstract}
A mean--field formulation is used to
investigate the bifurcation diagram for globally coupled tent maps by
means of an analytical approach. It is shown that the period
doubling sequence of the single site map induces a continuous 
family of periodic states in the coupled system. This type
of collective motion breaks the ergodicity of the coupled map
lattice. The stability analysis suggests that these states are stable 
for weak coupling strength but opens the possibility for more complicated 
types of motion in the regime of moderate coupling.   
\end{abstract}
\hspace*{0.6cm} 
\parbox{14cm}{
\begin{tabular}{ll}
PACS No.: & 05.45\\
Keywords: &\parbox{11cm}{Coupled map lattice, Synchronization
                         }\\
Running title: &  Globally Coupled Chaotic Systems\\
Submitted to: & Phys.~Rep.
\end{tabular}
}\\[0.5cm]
%
%
\section{Introduction}
Despite the deep insight which has been gained into the dynamics of low
dimensional chaotic systems and the sophisticated methods that have been 
developed for its analysis (cf.\cite{EcRu85A,GuHo86A}) there is a 
tremendous lack in the understanding of the influence of irregular
motion on high dimensional dynamics. The failure is twofold. On the one hand
it is not quite clear what kind of quantities are suitable for the
characterization of chaotic motion in extended systems. On the other hand 
there is no common sense what kind of model systems capture the relevant 
features of chaotic motion in
high dimensional phase spaces. Concerning the second problem systems of 
coupled maps have been proposed as suitable models \cite{Kane89A} because 
of the simplicity of numerical simulations and of the fact that maps have
proven as reasonable models for the investigation of low
dimensional chaos. Although there does not exists a satisfactory derivation
of coupled map lattices
form physically sound equations of motion a rough
estimate suggests \cite{YaFu84A} that they capture important
features of realistic systems\footnote{Contrary to some suggestions 
which can be found in the literature the coupling does not derive from a 
discrete Laplacian but from the propagator of the extended system.}.  
In addition to a pure numerical simulation, analytical and
even rigorous approaches have been applied successfully to simple
coupled map lattices. Especially it has been 
shown that for rapidly decreasing interaction between the different 
lattices sites the hyperbolic property of the local dynamics is
inherited to the coupled system, so that the space--time correlations
decrease exponentially \cite{BuSi88A,BrKu95A}. 
This mixing property is sometimes
used as an operating definition for space--time chaos although it does not
seem to be fully satisfactory. It has been suggested that a breakdown
of this regime can be understood in terms of phase transitions in the
two dimensional spin lattice which serves as a symbolic 
dynamics \cite{Buni95A}.

In order to contribute by means of an analytical approach
to the problem how chaos influences the motion
in high dimensional phase spaces 
I will consider in this article a rather simple but nontrival coupled
map lattice. For that purpose tent maps are chosen, as they have nontrivial 
features concerning their
bifurcation scenario, their symbolic dynamics and their spectral properties 
but allow for an analytical treatment (cf.~\cite{Doer85A}). For the coupling 
mechanism an all to all interaction is the simplest choice so that one ends up
with
\bea{aa}
x_{n+1}^{(\nu)} &=& \left(1-\eps\right) f\left(x_n^{(\nu)}\right)
+ \eps h_n =: T_n\left(x_n^{(\nu)}\right)\label{aaa}\\
h_n &:=& \frac{1}{N} \sum_{\mu=0}^{N-1} f\left(x_n^{(\mu)}\right)\label{aab}\\
f(x) &:=& 1 - a |x| \label{aac}\quad .
\eea
Here $N$ denotes the system size and we will dwell on the limit of
large system size in the sequel. A superficial inspection of the central
limit theorem together with the chaotic properties of the local
dynamics would suggest that global quantities like the coupling field
$h_n$ possess fluctuations which decrease as $N^{-1/2}$ with the
system size. But such a property requires the factorization of
the spatial correlation function at different sites (cf.~\cite{DiWi93A})
and is in this sense equivalent to the mixing property referred above.
It  may be violated due to the infinite range of 
coupling\footnote{Sometimes the
slightly misleading expression  ''Violation of the law of large numbers''
is used for this phenomenon in the literature.}, so that global quantities
obey a nontrivial dynamics even in the limit of infinite system 
size \cite{PiKu94A}. It has been suggested that the occurence of such a
kind of motion, which corresponds to a partial synchronization in the
extended system, is related to the nonhyperbolic properties of
the local dynamics \cite{Just95A}. For the simple model system
(\ref{aaa})--(\ref{aac}) I want to investigate how such a kind of motion
occurs.

For that purpose a mean--field like approach is used \cite{Kane92B}.
To keep the presentation self--contained and to set some notations
this formulation will be recapitulated in section 2. With its help the
bifurcation behaviour of periodic densities is analysed in section 3.
The highly nontrivial aspect of the stability of these solutions is
addressed in section 4.  
\section{Mean--field formulation}
The value of the coupling field (\ref{aab}) is solely determined by the
''one--site'' density
\be{ba}
\rho_n(x) := \frac{1}{N} \sum_{\mu=0}^{N-1} \delta\left(x-x_n^{(\mu)}
\right)
\ee
by dint of the relation
\be{bb}
h_n=\int f(x) \rho_n(x)\, dx \quad.
\ee
If one recalls that the dynamics is governed
by the effective map $T_n$ then the time evolution of the
density (\ref{ba}) is given by
\be{bc}
\rho_{n+1}(x) = \int \delta\left(x- T_n(y)\right) \rho_n(y)\, dy
=: {\cal L}_n \rho_n(x)
\ee
where the map $T_n$ depends itself  
on the density (cf.~eqs.(\ref{aaa}), (\ref{bb})). 
In this respect eq.(\ref{bc}) constitutes a
nonlinear evolution equation for the ''one--site'' density
(\ref{ba}). It should be stressed that this expression
is an exact consequence of the full evolution
equation (\ref{aaa}) as long as the density takes the form (\ref{ba}).
In additon the system size does not enter the formulation
in an explicit way.

It is supposed that the limit of infinite system size can be
performed in such a way that instead of the special form (\ref{ba})
every integrable density is accessible during the evolution.
Although this limit, understood in the weak sense, 
seems to be difficult to justify on a rigorous basis
this assumption is highly plausible on physical grounds and can be
supported by numerical simulations. By this assumption those
transients which increase with the system size are turned 
into stable solutions of the mean--field description\footnote{It is a
simple task to demonstrate this
property on the example of globally coupled shift maps.}. Hence the
stationary solutions of eq.(\ref{bc}) capture also the transient
behaviour which becomes quasi--stationary in the limit of large system size
and which governs the dynamics of large systems on accessible time scales.

The density (\ref{ba}) allows for the computation of any globally
averaged quantity. In addition it has been demonstrated \cite{ErPo95A} 
that in cases where the the mean--field dynamics (\ref{bc})
converges to a stable solution, the correlation properties between
different sites can be obtained from the ''one--site'' density also.
\section{Periodic Densities}
We are going to construct periodic solutions
$\rho_0,\ldots,\rho_{p-1}$, $\rho_p=\rho_0$ of the evolution
equation (\ref{bc}) by generalizing some ideas of \cite{Just95B}.
For reasons that will become obvious in the sequel
we consider periods $p=2^M$.
Eq.(\ref{ba}) tells us that $\rho_0$ is determined by
an invariant density of the map 
\be{ca}
\left(T_{p-1}\circ \ldots \circ T_0\right)(x) =
\eta_{p-1} - a_{red}|\eta_{p-2}- \ldots - a_{red}| \eta_0
- a_{red}|x|| \ldots | \quad .
\ee
Here the abbreviations 
\be{cb}
a_{red}:=(1-\eps) a, \quad
\eta_n : = 1-\eps a \int |x| \rho_n(x) \, dx
\ee
have been introduced. $\eta_n$ is positive as long as the coupling 
strength satisfies $\eps<1$. We further note the trivial fact that
every mean--field map $T_i(x)=\eta_i - a_{red} |x|$ is topological 
equivalent to a single tent map with a reduced parameter
\be{cba}
F(z)= 1-a_{red} |z| 
\ee
by a linear coordinate transformation $x=\eta_i z$.

Let us first consider the simple case that all parameters (\ref{cb}) are
equal $\eta_0=\ldots=\eta_p=\eta_*$. The $p^{th.}$ iterate of the map
(\ref{cba}) admits of $p$ different ergodic components for 
$\sqrt[2p]{2}< a_{red}<\sqrt[p]{2}$. They 
correspond to the $p$--band chaos in the tent map.
Let $\nu_i(x), i=0,\ldots,p-1$ denote the different normalized
ergodic absolutely continuous distributions and let $\nu_0$ 
denote that ergodic component which contains the critical point $z=0$. 
They obey the 
Frobenius--Perron equation
which reads $ \nu_{i+1}(x/\eta_*) = {\cal L}_* \nu_i(x/\eta_*)$. 
For $1\le i \le p-1$ it just leads to an affine transformation
of the distributions $\nu_i$. The density 
\be{cc}
\rho_0(x) := \sum_{i=0}^{p-1} \frac{1}{2^M} \frac{1}{\eta_*} \nu_i
\left(\frac{x}{\eta_*}\right)\quad .
\ee
constitutes by construction
a $p$ periodic orbit of eq.(\ref{ba}) which in fact is a fixed 
point. Finally the free parameter $\eta_*$ is determined by the relation 
(\ref{cb})
\be{cd}
\eta_* = 1- \eps \eta_* \sum_{i=0}^{p-1} \frac{1}{2^M} \int
|z| \nu_i(z)\, dz
\ee
where it is worth to mention again that the distributions $\nu_i$ depend only
on $a_{red}$ by definition. The meaning of this fixed point solution is quite
simple. Every periodic chaotic band contributes to the density (\ref{cc})
in such a way that the time dependencies cancel each other exactly.
Finally let us denote for later reference by $\sigma_i$ the symbol sequence
of the critical point that means $\sigma_{i}=\sgn(T^{i+1}_*(0))$.

\begin{figure}
\epsffile{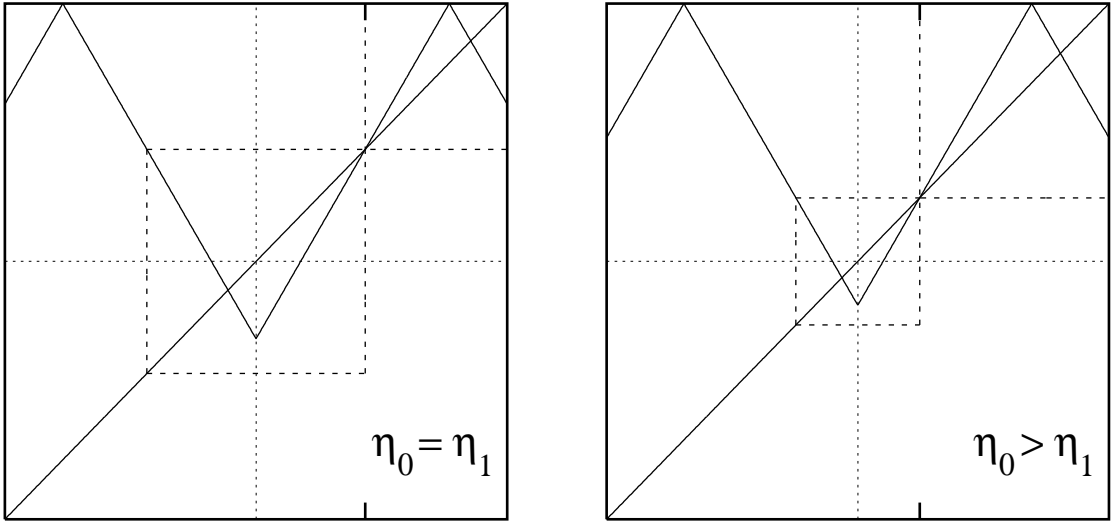}
\caption[ ]{\small Diagrammatic view of the second iterate $T_1\circ T_0$
for $\sqrt[4]{2} < a_{red} < \sqrt{2}$ and two different choices of
parameters $\eta_i$.\label{fig1}}
\end{figure}

We continue our construction of a continuous family of periodic solutions by
giving the different ergodic components different weight factors. For that
purpose one has to show that the components are not destroyed if the
parameters $\eta_n$ are not equal. In fact as sketched in Fig.~\ref{fig1} 
the change of these quantities will lead only to an affine transformation 
of the different components but not to topological changes.

To make this statement explicit consider a $p$ periodic
sequence of parameters $\eta_i$ 
close to the value $\eta_*$ mentioned above. Fix
a neighbourhood of the critical point $x=0$ that
contains the support of the former ergodic component. As the finite
itinerary of the critical point $\sigma_0,\ldots,\sigma_p$ is
unaltered for $\eta_i$ sufficiently close to $\eta_*$ the iterated
map (\ref{ca}) on this neighbourhood reads
\be{ce}
\left(T_{p-1}\circ \ldots \circ T_0\right)(x) =
\sum_{l=0}^{p-1} (-a_{red})^{p-l-1} \cdot \prod_{k=l}^{p-2}\sigma_k\cdot
\eta_l + (-a_{red})^{p-1} \cdot \prod_{k=0}^{p-2}\sigma_k \cdot |x| \quad .
\ee
Since this expression is in a
neighbourhood of the critical point
topological equivalent to
the $p^{th}$ iterate of the single tent map (\ref{cba}),  
eq.(\ref{ce}) admits of an
ergodic component $\tilde{\nu}_0(x)$. It is given in terms of
the corresponding ergodic component $\nu_0$ of the single map
by
\be{cea}
\tilde{\nu}_0(x)= \frac{1}{\Gamma}\nu_0\left(\frac{x}{\Gamma}\right)
\ee
where the scaling factor $\Gamma$ follows from the linear conjugacy 
to be\footnote{As long as the itinerant is not periodic the sum on
the left hand side does not vanish.}
\be{ceb}
\Gamma \sum_{l=0}^{p-1} (-a_{red})^{p-l-1} \cdot \prod_{k=l}^{p-2}\sigma_k
= \sum_{l=0}^{p-1} (-a_{red})^{p-l-1}\cdot \prod_{k=l}^{p-2} \sigma_k
\cdot \eta_l \quad.
\ee
By construction the sequence of densities
$\tilde{\nu}_{i+1}:= {\cal L}_i \tilde{\nu}_i$ is periodic with period $p$.
In view of the fact that the mean--field map $T_i$ is conjugate to the
single tent
map (\ref{cba}), and that the former acts as an affine
transformation for $\eta_i$ sufficiently close to $\eta_*$, it is
a simple task to express the densites $\tilde{\nu}_i$ in terms of the
ergodic components $\nu_i$ of the single map
\bea{cec}
\tilde{\nu}_i(x) &=& \frac{1}{\Gamma} \nu_i\left(\frac{x-\delta_i}
{\Gamma}\right)\label{ceca}\\
\delta_i &:=& \sum_{l=0}^{i-1} (-a_{red})^{i-l-1}\cdot \prod_{k=l}^{i-2}
\sigma_k \cdot
\left(\eta_l - \Gamma \right) \label{cecb} \quad .
\eea
Eq.(\ref{ceca}) just states that the densities $\tilde{\nu}_i$ follow
by an affine transformation from the ergodic components $\nu_i$. In order
that this representation is valid it is necessary, that the support of
these densities is not shifted across the critical point, that means
$\delta_i$ has to be sufficiently small. 
But from eqs.(\ref{ceb}) and (\ref{cecb})
it is obvious that this property holds for $\eta_i$ sufficiently close to
$\eta_*$. Now by construction the $p$ periodic convex sum
\be{cf}
\rho_n(x)=\sum_{i=0}^{p-1} \alpha_{n+i} \tilde{\nu}_i(x),\quad
\alpha_{n+p}=\alpha_n\in[0,1], \quad \sum_{k=0}^{p-1} \alpha_k=1 
\ee
satisfies eq.(\ref{bc}). It remains to show that
the weights $\alpha_i$ can be chosen in such a way that
the self--consistency condition (\ref{cb}) is valid. 
For that purpose one inserts the density
(\ref{cf}) into eq.(\ref{cb}) and obtains
\be{cg}
\eta_n=1-\eps a \sum_{i=0}^{p-1}\alpha_{n+i}
\left(\Gamma \zeta_i + \sigma_{i-1} \delta_i\right) \quad .
\ee
Here we have expressed the integrals in terms of the densities $\nu_i$
of the single map and have used the abbreviation
\be{ch}
\zeta_i := \int |z| \nu_i(z)\, dz \quad.
\ee
But now eqs.(\ref{cg}) and the definitions (\ref{ceb}) and (\ref{cecb})
yield a system of $p$ linear equations for the $p$ coefficients 
$\eta_n$ if one considers the weights $\alpha_i$ as given. The 
solution of this system completes the construction of the periodic 
density\footnote{One might argue, 
that the system is singular for a large set of parameter
values $(a,\eps)$. But as for fixed value of $a_{red}$ the
coefficients depend on the parameter $\eps a$ in a linear way
the determinant vanishes at most at $p$ isolated points.}. It is 
completely given in terms of the densities of the single map (\ref{cba}).

\begin{figure}
\epsffile{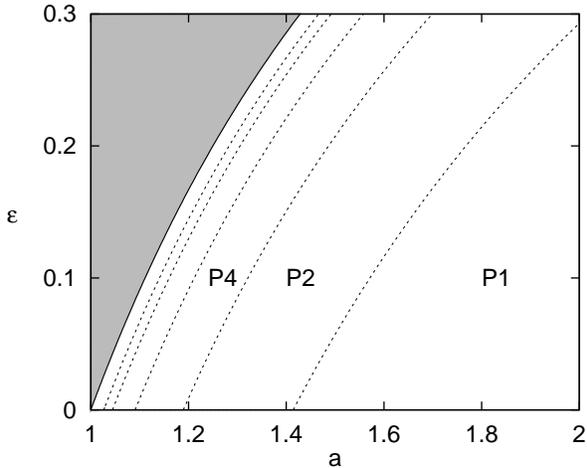}
\caption[ ]{\small Bifurcation diagram for globally coupled tent maps.
$Pn$ denotes the region where a continuous family of $n$ periodic
densities exists. The shaded area indicates the region where 
asymptotic periodicity gets lost. \label{fig2}}
\end{figure}

Hence one has found a continuous family of $p=2^M$ periodic solutions for
$\sqrt[2p]{2}<a_{red}<\sqrt[p]{2}$ which emerges by attributing 
slightly different weights to
the different ergodic components of the iterated mean--field map
(cf.~eq.(\ref{cf})). Fig.~\ref{fig2}
summarizes this result in a partial bifurcation diagram.
The period doubling of chaotic attractors of the single map translates into
the bifurcation of periodic densities of the coupled system. At each 
bifurcation line the period $p$ family of densities splits into a
period $2p$ family of densities. As was already mentioned in \cite{LMMa95A} at
$a_{red}=1$ asymptotic periodicty gets lost. In addition the
bifurcation at $a_{red}=\sqrt{2}$ has been observed in numerical simulations
some time ago \cite{Kane92B}.

\begin{figure}
\epsffile{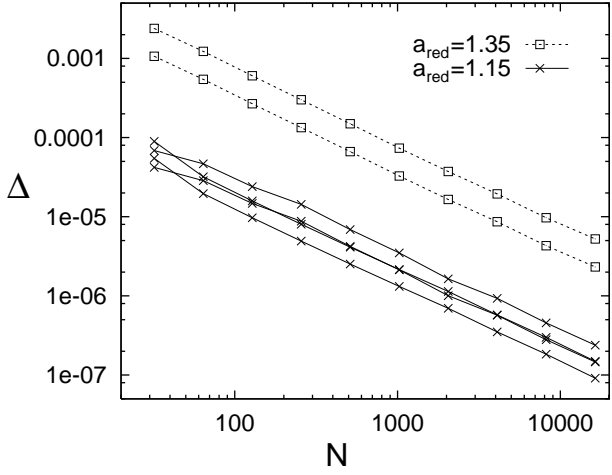}
\caption[ ]{\small Mean square deviation $\Delta$
of the second respectively fourth
iterate of the coupling field $h_n$ for
$\eps=0.1$ and two values of $a_{red}$ in dependence on the system size.
\label{fig3}}
\end{figure}

I have yet not claimed that the periodic densities are stable. One might argue
that on physical grouds these solutions are stable, as the ergodic components
are preserved in the coupled map lattice. But as was already mentioned in
\cite{ErPo95A} the problem of stability turns out to be
highly nontrivial. Furthermore in numerical simulations of finite
lattices one observes additonal oscillations \cite{Kane95A} which
cannot be attributed to the period doubling scenario.
Before I turn to the discussion of the
stability let me demonstrate by means of a numerical approach
that the densities might be stable at least for small coupling.
For that purpose simulations of eqs.(\ref{aaa})--(\ref{aac}) 
have been performed for
parameter values within the period 2 and 4 window. For different lattice
sizes one finds a corresponding
periodic motion in the coupling field on which a
''noisy'' component is superimposed. In order to check whether this
component is a finite size effect the mean square deviation, that means
the second cumulant of the $p^{th}$ iterate of $h_n$, has been analysed
in dependence on the system size (cf.~Fig.\ref{fig3}).
The behaviour according to the central limit theorem suggests 
that the nonperiodic component is a finite size effect and that these
periodic states are dynamically stable. But the numerical
values of the periodic density depend on the initial condition.
Hence the latter determines which member of the continous family of
periodic densities is selected as the stationary state.
\section{Stability Analysis}
The linear stability analysis of the equation of motion
(\ref{bc}) faces the problem that the densities are in general not
smooth functions. Following the ideas of \cite{ErPo95A}
this equation takes a somewhat simpler form if one restricts the analysis
to the case of densities which are
made of a countable infinite number of step functions
\be{da}
\rho_n(x) = \sum_{j=0} b_n^j \Theta\left(c_n^j -x\right) \quad .
\ee
Due to the piecewise linear shape of the tent map
this form is preserved during the evolution. Then
eqs.(\ref{bb}), (\ref{bc}) read (cf.~\cite{Mori96A})
\bea{db}
b_{n+1}^0 &=& \frac{1}{a_{red}} \sum_{j=0} b_n^j \left[ \sgn(c_n^j) +1
\right]\label{dba}\\
b_{n+1}^{j+1} &=& -\frac{1}{a_{red}} b_n^j \sgn(c_n^j)\label{dbb}\\
c_{n+1}^0 &=&  T_n(0)\label{dbc}\\
c_{n+1}^{j+1}&=&T_n\left( c_n^j\right) \label{dbd}\\
T_n(x) &=& 1-\eps+\eps h_n - a_{red} |x| \label{dbe}\\
h_n &=& 1 -\frac{a}{2} \sum_{j=0} b_n^j c_n^j | c_n^j| \label{dbf} \quad .
\eea

We will consider only the stability properties of the
fixed point solution (\ref{cc}) in the sequel. 
In terms of the representation (\ref{da})
this solution is obtained from eqs.(\ref{dba})--(\ref{dbd})
\bea{dc}
b_*^j &=& \left(-\frac{1}{a_{red}}\right)^j b_*^0 \prod_{k=0}^{j-1} 
\sigma_k\label{dca}\\
c_*^j &=& T_*^{j+1}(0) \label{dcb}\\
1 &=& b_*^0 \sum_{j=0} \left(-\frac{1}{a_{red}}\right)^j c_*^j 
\prod_{k=0}^{j-1} \sigma_k\label{dcc}
\eea
where again $\sigma_j:=\sgn(c_*^j)$ denotes the symbol sequence of the
critical point. To investigate the stability properties
of this solution let us consider
the eigenvalue problem that emerges from the formal linearization of the
evolution equations (\ref{dba})--(\ref{dbf})
\bea{dd}
\lambda \beta^0 &=& \frac{1}{a_{red}} \sum_{j=0} \beta^j (\sigma_j+1)
\label{dda}\\
\lambda \beta^{j+1} &=& -\frac{1}{a_{red}} \beta^j \sigma_j \label{ddb}\\
\lambda \gamma^0 &=& \eps \delta h\label{ddc}\\
\lambda \gamma^{j+1} &=& \eps \delta h - a_{red} \gamma^j \sigma_j
\label{ddd}\\
\delta h &:=& -\frac{a}{2} \sum_{j=0} \left( c_*^j |c_*^j| \beta^j
+ 2 b_*^j |c_*^j| \gamma^j\right) \label{dde} \quad.
\eea
Since these equations decouple the eigenvalue problem can be analysed
on both invariant subspaces separately.

With the solution of eq.(\ref{ddb})
\be{de}
\beta^j = \left(-\frac{1}{a_{red}\lambda}\right)^j \beta^0 \prod_{k=0}^{j-1}
\sigma_k
\ee
eq.(\ref{dda}) yields the characteristic equation
\be{df}
0 = \left(1- z\right) \sum_{j=0} (-z)^j \prod_{k=0}^{j-1} \sigma_k,
\quad z:=\frac{1}{a_{red}\lambda}\quad .
\ee
The sum on the right hand side is just the $\lambda$--expansion and
yields the spectrum of the Frobenius--Perron operator of the fixed 
point map $T_*$ \cite{Doer85A}. Hence the
spectrum is contained within the unit circle and the eigendirections belong 
to the center--stable manifold of the fixed point density. Obviously
there occurs a doubly degenerated eigenvalue $\lambda=1$. It is 
caused by two constants
of motion $\sum_j b_n^j=0$, $\sum_j b_n^j c_n^j=1$ of 
eqs.(\ref{dba})--(\ref{dbf})
which reflect the normalization of the density and the boundary condition
$\rho_n(-1)=0$.

Using the solution of eqs.(\ref{ddc}), (\ref{ddd})
\be{dg}
\gamma^j=\sum_{l=0}^j \prod_{k=1}^{l}\left(-\frac{a_{red}}{\lambda} 
\sigma_{j-k} \right) \cdot \frac{\eps \delta h}{\lambda}
\ee
eq.(\ref{dde}) yields for $\beta^j\equiv 0$
the characteristic equation
\be{dh}
1=-\frac{a \eps}{\lambda} \sum_{l=0} \left(-\frac{a_{red}}{\lambda}\right)^l
\sum_{j=l} b_*^j | c_*^j| \prod_{k=1}^{l} \sigma^{j-k}\quad.
\ee
If one takes the representation (\ref{dca}) of the fixed point into
account one ends up with
\be{di}
1 = -\frac{a \eps}{\lambda} \sum_{l=0} \kappa_l
\left(-\frac{1}{\lambda}\right)^l
\ee
where the coefficients are given by
\be{dj}
\kappa_l := b_*^0 \sum_{j=0} \left(-\frac{1}{a_{red}}
\right)^j |c_*^{j+l}| \prod_{k=0}^{j-1} \sigma_k \quad .
\ee
Although one has obtained a quite simple expression which can be
evaluated by means of a numerical 
approach\footnote{The crucial step in this approach is the desired truncation
of the potentially infinite series.} it seems to be difficult to 
make general statements about the solutions of eq.(\ref{dj}) respectively its
analytical continuation. But if one restricts to the simple case
that the itinerary terminates in the fixed point of the map $T_*$,
that means $|c_*^l|=c_F$, $l\ge l_0$, then the coefficients obey
\be{dk}
\alpha_l=b_*^0 c_F 
\sum_{j=0} \left(-\frac{1}{a_{red}}
\right)^j \prod_{k=0}^{j-1} \sigma^k=0, \quad l\ge l_0\quad .
\ee
To establish this identity one has used the fact that the sum represents the
$\lambda$--expansion of the fixed point map $T_*$ and that its
Frobenius--Perron operator
admits of the eigenvalue $\lambda=1$ \cite{Doer85A}. 
Hence the characteristic equation 
(\ref{di}) reduces to a polynomial which yields for sufficiently small
coupling only eigenvalues within the unit circle.
\section{Discussion}
The bifurcation diagram depicted in Fig.~\ref{fig2} summarizes main 
results of this article. It provides a foliation of the region of asymptotic
periodicity $a_{red}>1$. In each region a continuous family of
periodic states, which is generated by the different ergodic components 
of the iterated tent map,
accounts for the time dependency of the stationary solution. 

The stability analysis suggests that these states are stable at least
for small coupling strength $\eps$. But there are several limitations in 
this analysis. For the stability of the fixed point density 
the estimates presented in section 4
are not uniformly valid in the
parameter space and the problem whether stability is attained on
a parameter set of sufficiently large measure is not solved.
However numerical simulations
and a numerical analysis of the eigenvalue equations (\ref{dda})--(\ref{dde})
\cite{Mori96A} suggest stabilty for small coupling. 
In addition one should recall that the discussion of stability 
presented above was based on a formal linearization of the 
evolution equation. Such a procedure takes deviations of the
parameters $c^j$ with equal weight into account. This approach seems to
overestimate the effect of these deviations on the density (\ref{da})
since the coefficients $b^j$ decay exponentially. Hence one has to
specify the space on which the eigenvalue problem has to be treated
which may restrict the spectrum to a few relevant eigenvalues
(cf. e.g.~\cite{SaHa92A} for such a phenomenon in the context of
the Frobenius--Perron operator). Despite these shortcommings the 
approach presented here might be a basis
for further investigations.

The periodic states investigated in this article explain in a clear 
fashion the time dependence of globally averaged quantities.
Therefore one mechanism for
time fluctuations in globally averaged 
quantities seems to be well understood. But numerical simulations
of coupled tent maps
show additional quasiperiodic solutions \cite{Kane95A}. The explanation of 
this kind of collective motion which resembles in some respect a 
Hopf bifurcation seems to be one challenge for the further research in 
this field. In addition it might be this kind of motion which 
appears in the frequently analysed case of coupled logistic maps.

The feasibility of analytical computations on piecewise linear
map lattices make these kind of models suitable for studying the
influence of chaotic motion in high dimensional systems. Even the
apparently simple model treated in this article shows several
unexpected features and it is far from being solved completely.
Other problems may be attacked by this kind of model also, e.g.~the
effect of a long but not infinte range of coupling in connection with
the validity of a mean--field treatment, 
or the construction of the symbolic dynamics
and the effect of pruning of symbol sequences. But these 
questions are left for future work.
\section*{Acknowledgement}
The author is indebted to the ''Deutsche Forschungsgemeinschaft'' 
and to the ''Ver\-ei\-ni\-gung der Freunde der TH--Darmstadt''
for financial support. This work was performed within
a program of the Sonderforschungsbereich 185 Darmstadt--Frankfurt, FRG.
%
%
%

\end{document}